\documentclass[%
 reprint,
superscriptaddress,
 amsmath,amssymb,
 aps,pre,hyphens,floatfix
]{revtex4-2}
\usepackage{amsfonts}
\usepackage{amsmath}
\usepackage[]{amsmath}
\usepackage{amssymb}
\usepackage{multirow} 
\usepackage{enumerate}
\usepackage[usenames,dvipsnames,svgnames,table]{xcolor}
\usepackage{IEEEtrantools}
\usepackage{graphicx}
\usepackage[normalem]{ulem}  

\begin{document}

\preprint{APS/123-QED}

\title{Stochastic dynamics generation of directed percolation using a two-channel convolutional neural network}
\author{Feng Gao}
\affiliation{Key Laboratory of Quark and Lepton Physics (MOE) and Institute of Particle Physics, Central China Normal University, Wuhan 430079, China}

\author{Jianmin Shen}
\affiliation{College of engineering and technology, Baoshan University, Baoshan 678000, China}
\affiliation{Key Laboratory of Quark and Lepton Physics (MOE) and Institute of Particle Physics, Central China Normal University, Wuhan 430079, China}

\author{Wei Li}
\email[]{liw@mail.ccnu.edu.cn}
\affiliation{Key Laboratory of Quark and Lepton Physics (MOE) and Institute of Particle Physics, Central China Normal University, Wuhan 430079, China}
\affiliation{SCIQ Lab, École Supérieure d'Informatique Électronique Automatique, Ivry-sur-Seine 94200, France}

\date{\today}

\begin{abstract}

Neural networks provide a promising approach to learning and generating the stochastic dynamics of nonequilibrium systems. In this study, we propose a stochastic dynamics generative method based on a two-channel convolutional neural network, using the (1+1)-dimensional directed bond percolation model to investigate its ability to learn local stochastic transition rules and reproduce stochastic dynamics across different system sizes. The proposed method takes the current particle configuration and the control parameter as inputs, learns the conditional activation probabilities at the next time step through local convolutions, and generates stochastic temporal evolution through iterative Bernoulli sampling. The results show that the network-generated dynamics agree well with Monte Carlo simulations across different control parameters. Furthermore, applying the trained network to a substantially larger system with $L=10000$ yields an estimated critical probability of $p_c=0.643$ and an effective density decay exponent of $\delta\approx0.1592$, close to the established theoretical values. These findings demonstrate that the proposed method can reproduce the stochastic dynamics of directed percolation and capture its macroscopic dynamical and statistical properties across different spatial sizes.


\end{abstract}

\maketitle

\section{Introduction}
\label{intro}

Nonequilibrium phase transitions~\cite{hinrichsen2006non,lubeck2004universal} are a central topic in statistical physics and occur in a broad range of systems governed by irreversible dynamics, including epidemic spreading~\cite{balcan2011phase}, chemical reactions~\cite{schlogl1972chemical,kumar2014polymorphic}, and population dynamics~\cite{davis2020phase,heffern2021phase}. Directed percolation (DP)~\cite{cardy1980directed,domany1981directed} serves as a paradigmatic model for investigating such transitions and defines one of the most extensively studied universality classes in nonequilibrium statistical physics~\cite{lubeck2004universal,domany1981directed}. Despite the simplicity of its microscopic spreading rules, DP exhibits complex collective behavior at the macroscopic level. Conventional approaches to studying DP include Monte Carlo (MC) simulations, mean-field theory, and renormalization-group methods~\cite{hinrichsen2006non,domany1981directed,lubeck2004universal}. Among these approaches, MC simulations remain an essential numerical tool for investigating low-dimensional DP systems, in which fluctuations play a crucial role.

In recent years, advances in machine learning have facilitated the application of neural networks~\cite{koch2018mutual,han2025graph,shen2022transfer,shen2021supervised,arce2020phase} to statistical physics, including the study of nonequilibrium phase transitions. These applications encompass a wide range of tasks, including phase classification~\cite{mobilia2007phase,woollam2019convolutional,dokht2019seismic}, critical point identification~\cite{gao2026tricritical,gao2025neural}, critical exponent estimation~\cite{intrator1991feature,li2019extracting}, physical feature extraction~\cite{shen2025learning,wang2026siamese}, and dynamical prediction~\cite{mohajerin2018multi,shen2024machine,efthymiou2019super,jiang2019model}. Despite these advances, many existing approaches primarily focus on extracting physical information from system configurations through classification, regression, or feature extraction, rather than explicitly learning the stochastic transition rules governing their evolution. In contrast, generative models~\cite{salakhutdinov2015learning,ruthotto2021introduction,oussidi2018deep,ranzato2011deep} can learn probability distributions from data and generate new configurations.

Generative approaches, including generative adversarial networks~\cite{liu2017simulating,huembeli2018identifying}, variational autoencoders~\cite{luchnikov2019variational,tang2024learning}, normalizing flows~\cite{kobyzev2020normalizing,zhai2024normalizing}, and diffusion models~\cite{cao2024survey,yang2023diffusion}, have been increasingly applied to statistical physics. Boltzmann machines trained on Ising configurations have been used to reproduce thermodynamic observables~\cite{torlai2016learning}, while convolutional generative adversarial networks have generated Ising configurations conditioned on temperature~\cite{liu2017simulating}. Variational autoencoders have also been employed to reconstruct quantum many-body states from measurement data~\cite{luchnikov2019variational}. Variational autoregressive networks enable direct sampling and explicit evaluation of normalized configuration probabilities~\cite{wu2019solving}. Flow-based approaches have been incorporated into Markov chain MC methods for lattice field theory~\cite{albergo2019flow}, while Boltzmann generators employ learned transformations to sample equilibrium states of many-body systems~\cite{noe2019boltzmann}. However, without explicitly incorporating temporal conditioning or dynamical constraints, these models do not necessarily guarantee that the generated configurations obey the underlying stochastic transition rules. Moreover, directly modeling complete system configurations may be constrained by input dimensionality, model complexity, and training data requirements, potentially limiting the applicability of trained networks to systems of different spatial sizes. Therefore, rather than learning the probability distribution of complete configurations, learning the conditional transition probabilities determined by the current configuration and control parameter offers an alternative approach to generating nonequilibrium stochastic dynamics that more closely reflects the underlying microscopic evolution mechanisms.

In this work, we propose a method for generating stochastic dynamics based on a two-channel convolutional neural network, using the (1+1)-dimensional directed bond percolation model to investigate the ability of neural networks to learn microscopic stochastic transition rules and reproduce macroscopic dynamical behavior in nonequilibrium systems. The proposed method takes the particle configuration at the current time step and the control parameter as two input channels. Through local convolutions, the network learns the conditional activation probability of each site at the next time step, from which the subsequent configuration is generated through Bernoulli sampling~\cite{chen1997statistical}. The generated configuration is then fed back into the network, enabling successive stochastic evolution without requiring additional MC-generated configurations during the generation process. Owing to its fully convolutional architecture, the learned local transition rules can be applied to systems of different spatial sizes using shared convolutional parameters, without retraining the network for each system size. We further apply the trained model to systems substantially larger than those used during training to investigate the long-time decay of particle density near criticality and analyze the associated critical dynamical behavior through power-law fitting. This study aims to explore the feasibility of generating nonequilibrium stochastic dynamics by learning local conditional transition probabilities, providing an alternative framework for investigating the microscopic evolution mechanisms and macroscopic statistical properties of complex stochastic systems using neural networks.

\section{model}

DP is a paradigmatic model for investigating nonequilibrium continuous phase transitions and is widely used to describe stochastic spreading processes with absorbing states, including contact processes, epidemic spreading, and catalytic reactions. In this study, we consider the (1+1)-dimensional directed bond percolation model, defined on a one-dimensional lattice of $L$ sites evolving in discrete time steps. Each site can occupy one of two states, denoted by $x_i^t\in\{0,1\}$, where $x_i^t$ represents the state of site $i$ at time step $t$, with $1$ and $0$ corresponding to the active and inactive states, respectively. Periodic boundary conditions are imposed, and the system is initialized in a fully occupied state.

The evolution of the directed bond percolation model follows local stochastic propagation rules. At each time step, an active site can transmit activity to its two neighboring sites at the next time step through directed bonds, each independently transmitting activity with probability $p$, which serves as the control parameter. A target site becomes active if at least one propagation event is successful; otherwise, it remains inactive. Accordingly, the state of site $i$ at time step $t+1$ is determined by the states of its two neighboring predecessor sites at time step $t$, according to

\begin{equation}
    x_i^{t+1}=
    \begin{cases}
    1, & \text{if } x_{i-1}^t=1 \text{ and } z_i^{-}<p, \\
    1, & \text{if } x_{i+1}^t=1 \text{ and } z_i^{+}<p, \\
    0, & \text{otherwise},
    \end{cases}
    \label{equ1}
\end{equation}

where $z_i^{-}$ and $z_i^{+}$ are independent random variables uniformly distributed over the interval $[0,1]$. Based on these propagation rules, the conditional activation probability of site $i$ at the next time step is expressed as

\begin{equation}
    q_i^{t+1}
    =1-(1-p)^{x_{i-1}^{t}+x_{i+1}^{t}}.
    \label{equq}
\end{equation}

When both predecessor sites are inactive, the activation probability of the target site is zero. When only one predecessor site is active, the activation probability is $p$, whereas when both predecessor sites are active, it increases to $2p-p^2$. Thus, the microscopic evolution of the system is governed by the local configuration and the control parameter. Once all sites become inactive, no further activity can be generated, and the system remains in this configuration, referred to as the absorbing state.

\begin{figure*}[!thb]
\centering
    \includegraphics[width=0.98\textwidth]{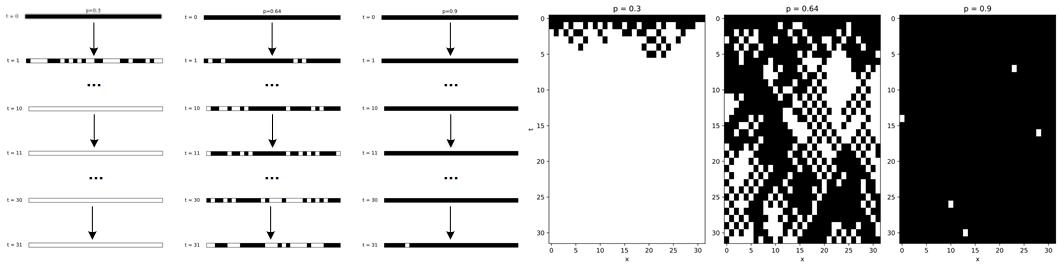}
\caption{Schematic illustration of the stochastic evolution of the (1+1)-dimensional directed bond percolation model at different propagation probabilities.}
\label{configuration}
\end{figure*}

As the propagation probability $p$ increases, the DP system undergoes a continuous phase transition~\cite{hinrichsen2006non} from the absorbing phase to the active phase at the critical point $p_c$. Fig~\ref{configuration} illustrates the stochastic evolution of the (1+1)-dimensional directed bond percolation model at $p=0.3$, $0.64$, and $0.9$. The left panels show spatial configurations at selected time steps, while the right panels present the corresponding spatiotemporal configurations, with black and white representing active and inactive sites, respectively. For $p=0.3$, the activity rapidly decays, and the system eventually reaches the absorbing state. At $p=0.64$, near the critical point, pronounced spatiotemporal fluctuations occur, and activity persists for a longer time. In contrast, for $p=0.9$, the system maintains a high density of active sites, exhibiting the characteristic behavior of the active phase.

To quantitatively characterize the temporal evolution of the system, the average particle density at time step $t$ is defined as

\begin{equation}
    \rho(t)=
    \left\langle
    \frac{1}{L}\sum_{i=1}^{L}x_i^t
    \right\rangle,
    \label{rho}
\end{equation}

At the critical point, the particle density exhibits an asymptotic power-law decay in sufficiently large systems~\cite{lubeck2004universal},

\begin{equation}
    \rho(t)\sim t^{-\delta},
    \label{delta}
\end{equation}

where $\delta$ is the critical exponent characterizing the decay of the particle density. By examining the temporal evolution of the particle density at different propagation probabilities, the absorbing-state phase transition and the associated critical dynamics can be investigated.

The stochastic evolution of the DP model is determined by local state transitions between successive time steps, providing a well-defined framework for learning microscopic stochastic dynamics using neural networks. In this study, pairs of configurations at consecutive time steps obtained from MC simulations are used to construct the training dataset. A convolutional neural network is then employed to learn the conditional activation probabilities associated with different local configurations and control parameters, enabling the generation of stochastic configurations and their iterative evolution over time.

\section{Neural Network based Approach}

Generative models have attracted considerable attention in recent years. These models typically learn the probability distribution of training data through latent variable mappings, adversarial training, or iterative denoising processes, enabling the generation of samples with similar statistical properties. However, they primarily focus on capturing the overall data distribution rather than explicitly learning the dynamical transition rules governing microscopic states. In nonequilibrium stochastic systems such as DP, spatiotemporal evolution is governed by local stochastic interactions and control parameters, from which macroscopic statistical properties and critical behavior emerge.

\begin{figure*}[!thb]
\centering
    \includegraphics[width=\linewidth,trim=185pt 200pt 90pt 230pt,clip]{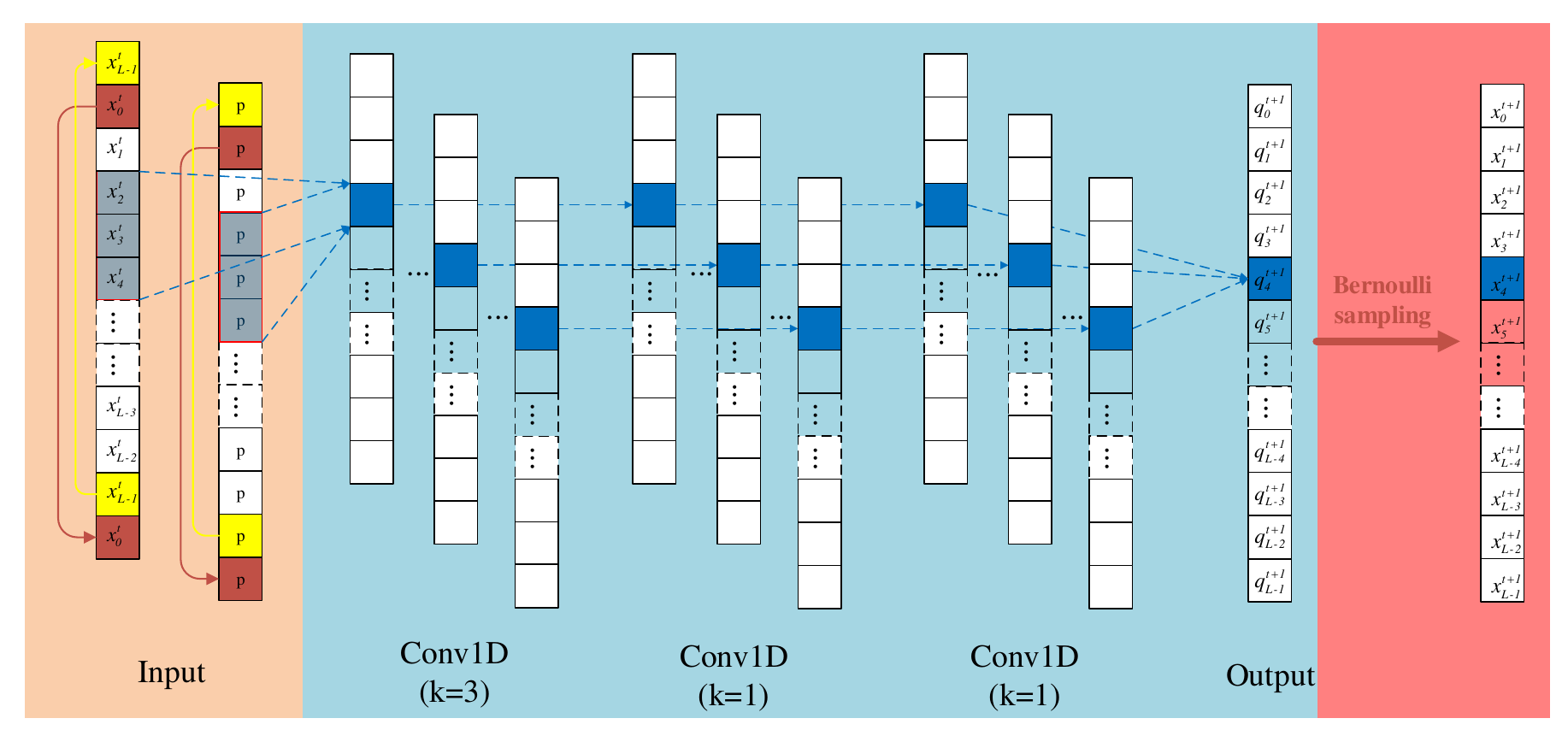}
    \caption{Schematic illustration of the two-channel convolutional neural network for one-step stochastic evolution. The input consists of the binary configuration $\boldsymbol{x}^{t}$ at time step $t$ and the control parameter $p$, which is constant across all spatial sites. Periodic padding is applied to preserve neighborhood information across the lattice boundaries. The network comprises three one-dimensional convolutional layers with kernel sizes of $3$, $1$, and $1$, and output channels of $64$, $64$, and $1$, respectively. The first two layers employ ReLU activation functions, while the output layer uses a Sigmoid activation function to predict the sitewise conditional activation probabilities $\boldsymbol{q}^{t+1}$. The subsequent binary configuration $\boldsymbol{x}^{t+1}$ is generated through independent Bernoulli sampling at each site and can be fed back into the network for iterative stochastic evolution.}
    \label{fig:dp_cnn}
\end{figure*}

In this study, we develop a stochastic dynamics generative model based on a two-channel convolutional neural network, as illustrated in Fig.~\ref{fig:dp_cnn}. The network takes the spatial configuration at the current time step and the control parameter as inputs and employs local convolutional operations to capture spatial dependencies between neighboring sites and learn the conditional activation probabilities at the next time step. The subsequent configuration is generated through independent Bernoulli sampling at each site. By iteratively feeding the generated configurations back into the network, the model enables the stochastic temporal evolution of the DP system. Owing to its fully convolutional architecture, the learned parameters can be applied to systems of different spatial sizes without requiring size-specific retraining.

The convolutional neural network architecture employed in this study is illustrated in Fig.~\ref{fig:dp_cnn}. The network comprises three one-dimensional convolutional layers and learns the mapping from the current spatial configuration to the activation probabilities of individual sites at the next time step, thereby modeling the local stochastic dynamics of the DP system. For a system of spatial size $L$, the configuration at time step $t$ is represented as $X^{t}=(x_1^t,x_2^t,\ldots,x_L^t)$, where $x_i^t\in\{0,1\}$ denotes the state of site $i$, with $0$ and $1$ corresponding to the inactive and active states, respectively. To incorporate the control parameter, $p$ is replicated across all spatial sites and combined with the current configuration to form a two-channel input of shape $L\times2$. The first channel contains the spatial configuration, while the second contains the control parameter, enabling the network to learn the stochastic evolution rules across different values of $p$ within a unified framework.

The first convolutional layer contains 64 filters with a kernel size of $k=3$ to extract local features from each site and its nearest neighbors. To maintain the periodic boundary conditions of the DP system, periodic padding is applied before convolution, ensuring that boundary sites have the same neighborhood structure as interior sites. The second convolutional layer employs 64 filters with a kernel size of $k=1$ to perform nonlinear transformations of the extracted features. Both layers use the rectified linear unit (ReLU) activation function. The output layer consists of a single filter with a kernel size of $k=1$, followed by a Sigmoid activation function, to predict the activation probability of each site at the next time step. The resulting output has a shape of $L\times1$, preserving the spatial size of the system.

Since only the first convolutional layer employs a kernel size of $k=3$, while the subsequent layers use $k=1$, the predicted activation probability at each site depends only on its current state, the states of its nearest neighbors, and the control parameter $p$. The resulting local mapping can be expressed as

\begin{equation}
    q_i^{t+1}=f_\theta\left(x_{i-1}^{t},x_i^{t},x_{i+1}^{t},p\right),
    \label{equq_ML}
\end{equation}

where $f_\theta$ denotes the nonlinear mapping parameterized by the trainable network parameters $\theta$, and $q_i^{t+1}$ represents the conditional activation probability of site $i$ at the next time step. Owing to the spatial sharing of convolutional parameters, the number of trainable parameters is independent of the system size, enabling the learned local transition rules to be applied to systems of different spatial sizes without retraining.

The network is trained in a supervised manner using pairs of consecutive configurations generated by MC simulations. The current configuration and the corresponding control parameter serve as inputs, while the configuration at the next time step is used as the training target. Since different local configurations in the DP system may occur with substantially different frequencies, a weighted binary cross-entropy (WBCE) loss function is employed to mitigate the imbalance in their contributions to the training process. As illustrated in Fig.~\ref{fig:local_structures}, a local structure consisting of a target site and its two nearest neighbors has $2^3=8$ possible binary configurations. To balance their contributions, weights are assigned inversely proportional to the occurrence frequencies of the corresponding local configurations. The WBCE loss function is defined as

\begin{equation}
\begin{aligned}
\mathcal{L}_{\mathrm{WBCE}}
=&-\frac{1}{NL}
\sum_{n=1}^{N}\sum_{i=1}^{L}
w_{i,n}
\Big[
x_{i,n}^{t+1}\ln q_{i,n}^{t+1}\\
&\quad+
(1-x_{i,n}^{t+1})
\ln(1-q_{i,n}^{t+1})
\Big],
\end{aligned}
\label{equL}
\end{equation}

where $N$ denotes the number of training samples, $L$ is the spatial size, and $w_{i,n}$ is the weight assigned to the local configuration centered at site $i$ in the $n$-th sample. For the $c$-th local configuration observed in the training dataset, the corresponding weight is defined as

\begin{equation}
    w_c=\frac{NL}{K n_c},
    \label{equ3}
\end{equation}

where $n_c$ represents the total number of occurrences of the $c$-th local configuration, and $K$ denotes the number of distinct local configurations observed in the training dataset. This weighting scheme assigns greater importance to less frequent local configurations, thereby reducing the influence of imbalanced occurrence frequencies during network training.

The network is trained using the Adam optimizer with an initial learning rate of $10^{-3}$, a batch size of 64, and a maximum of 100 epochs. During training, the learning rate is adaptively reduced based on the training loss, and early stopping is employed to prevent unnecessary training iterations. The network parameters corresponding to the lowest training loss are retained for subsequent analysis.

Once trained, the network predicts the sitewise activation probabilities at the next time step, denoted by $\boldsymbol{q}^{t+1}=(q_1^{t+1},q_2^{t+1},\ldots,q_L^{t+1})$, from the current configuration and the control parameter. To account for the stochastic nature of DP dynamics, the state of each site at the next time step is independently sampled from a Bernoulli distribution with the corresponding predicted activation probability:

\begin{equation}
    x_i^{t+1}\sim
    \operatorname{Bernoulli}(q_i^{t+1}).
    \label{equ4}
\end{equation}

The resulting binary configuration $\boldsymbol{x}^{t+1}$ is then combined with the same control parameter and fed back into the network. By iteratively applying the probability prediction and Bernoulli sampling procedures, the model generates stochastic evolution trajectories from a given initial configuration. Unlike deterministic configuration prediction, this approach explicitly incorporates stochastic sampling based on learned local transition probabilities, enabling different realizations to emerge from identical initial conditions. The resulting framework provides a neural-network-based approach to investigating the dynamical and statistical properties of the DP system.

\section{Result}

To evaluate the ability of the two-channel convolutional neural network to learn the local stochastic transition rules of directed bond percolation, we first examine the predicted one-step activation probabilities for different local configurations. Since only the first convolutional layer uses a kernel size of $k=3$, while the subsequent layers employ $k=1$, the predicted activation probability at each site depends only on its current state, the states of its two nearest neighbors, and the control parameter $p$. 

\begin{figure}[!thb]
    \centering
    \includegraphics[width=1\linewidth,trim=0 400pt 340pt 0pt,clip]{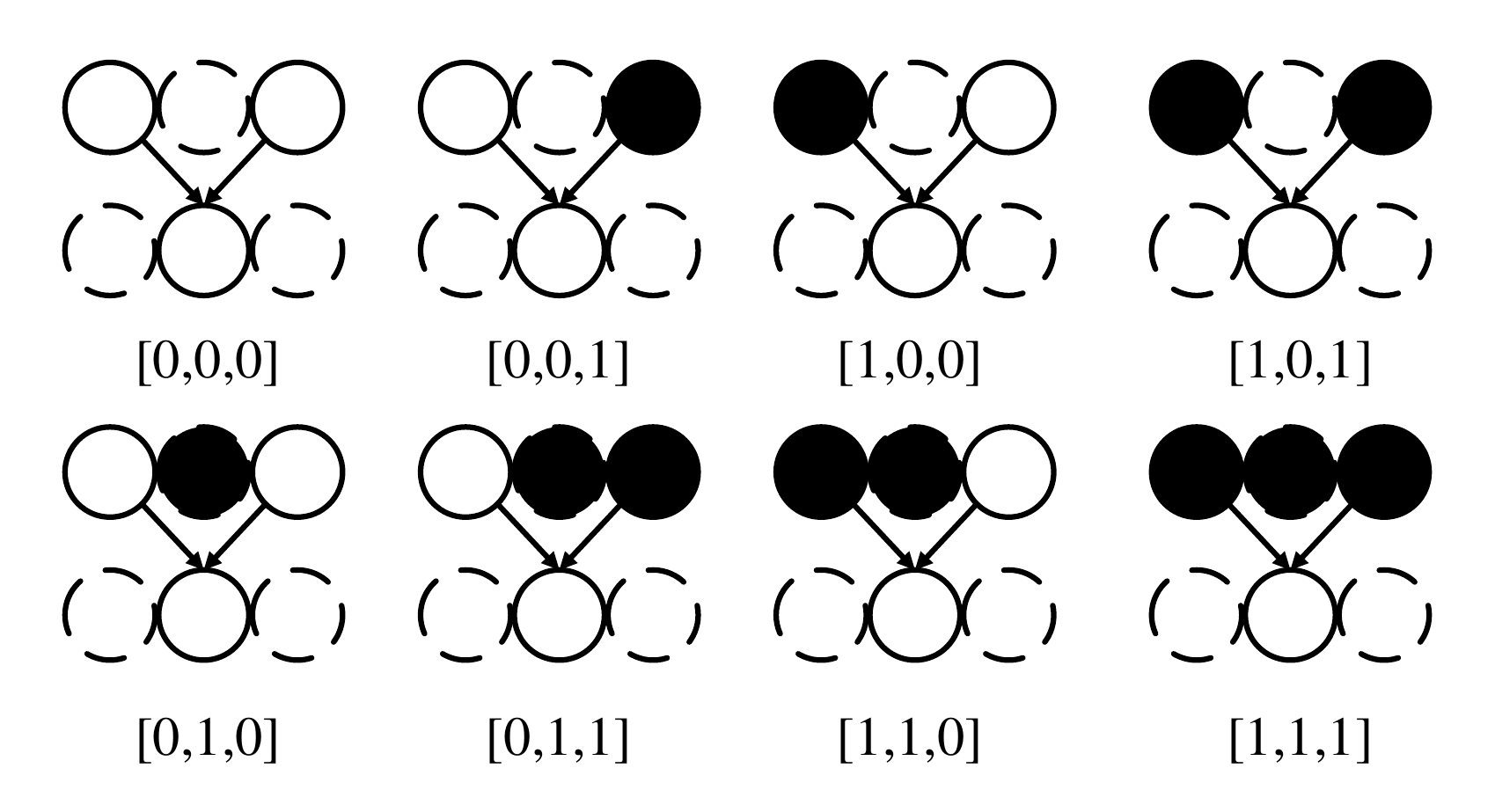}
    \caption{Schematic illustration of the eight three-site local configurations, represented as $[0,0,0]$, $[0,0,1]$, $[1,0,0]$, $[1,0,1]$, $[0,1,0]$, $[0,1,1]$, $[1,1,0]$, and $[1,1,1]$, where the three entries correspond to the left neighboring site, central site, and right neighboring site, respectively. Filled black circles and open circles denote active and inactive sites, respectively.}
    \label{fig:local_structures}
\end{figure}

These three sites form $2^3=8$ possible local configurations, as illustrated in Fig.~\ref{fig:local_structures}. Filled and open circles represent active and inactive sites, respectively. The three upper sites represent the input configuration at the current time step, while the lower central site denotes the target site at the next time step.

According to the stochastic propagation rules of directed bond percolation, the theoretical activation probability depends only on the number of active predecessor sites and is independent of the current state of the central site. For zero, one, or two active predecessors, the corresponding activation probabilities are $0$, $p$, and $2p-p^2$, respectively. Consequently, the eight local configurations can be classified into three groups according to their theoretical activation probabilities: $[0,0,0]$ and $[0,1,0]$ correspond to $q=0$; $[0,0,1]$, $[1,0,0]$, $[0,1,1]$, and $[1,1,0]$ correspond to $q=p$; and $[1,0,1]$ and $[1,1,1]$ correspond to $q=2p-p^2$.

\begin{figure*}[!thb]
    \centering
    \includegraphics[width=0.98\linewidth]{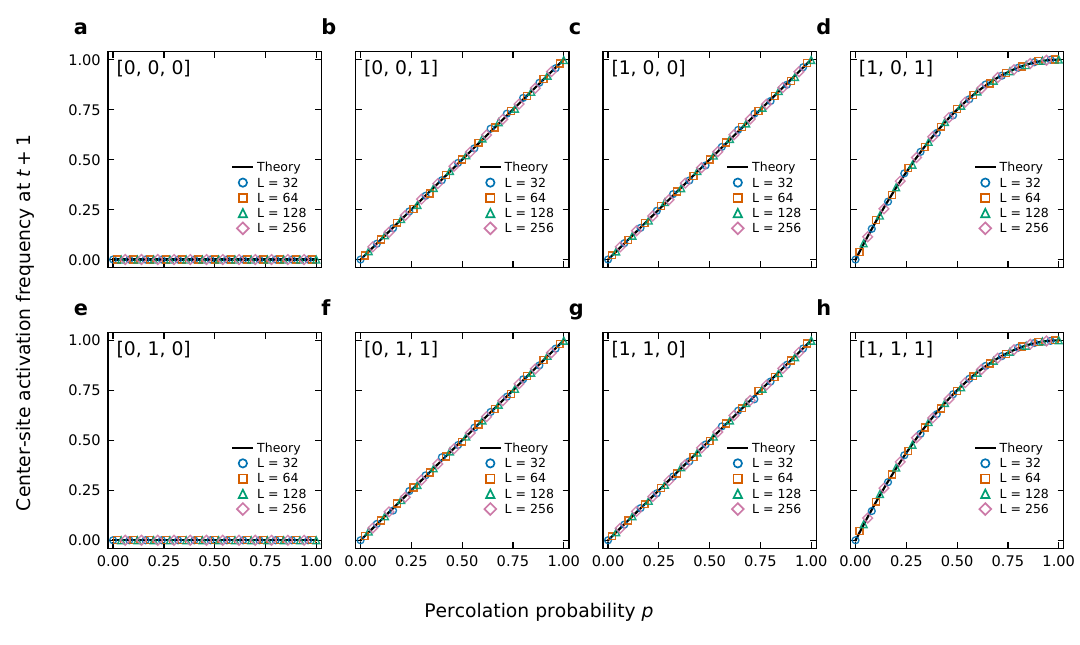}
    \caption{One-step activation probabilities for eight local configurations across different system sizes. Panels (a)--(h) correspond to the configurations $[0,0,0]$, $[0,0,1]$, $[1,0,0]$, $[1,0,1]$, $[0,1,0]$, $[0,1,1]$, $[1,1,0]$, and $[1,1,1]$, respectively. The horizontal axis represents the propagation probability $p$, and the vertical axis shows the activation probability of the target site at time step $t+1$, estimated from Bernoulli-sampled configurations. For each value of $p$ and system size, $2000$ independent random initial configurations are generated, with each site initially active with a probability of $1/2$. Blue circles, orange squares, green triangles, and purple diamonds represent the results for $L=32$, $64$, $128$, and $256$, respectively, while the black solid curves indicate the theoretical activation probabilities. All predictions are obtained using the same network trained on the $L=32$ system, with periodic boundary conditions.}
    \label{fig:multisize_transfer}
\end{figure*}

To evaluate the accuracy of the learned local transition probabilities and their applicability across different spatial sizes, we apply the network trained on a system with $L=32$ to systems with $L=32$, $64$, $128$, and $256$. For each system size, $51$ equally spaced propagation probabilities are selected over the interval $p\in[0,1]$. At each value of $p$, $2000$ independent random initial configurations are generated, with each site initially active with a probability of $1/2$. The configurations and their corresponding control parameters are combined to form two-channel inputs, from which the network predicts the sitewise activation probabilities at the next time step. The subsequent configurations are generated through independent Bernoulli sampling, and the activation probabilities for each local structure are estimated from the resulting activation frequencies. The same trained network and periodic boundary conditions are used for all system sizes without retraining.

Fig~\ref{fig:multisize_transfer} compares the estimated one-step activation probabilities for the eight local configurations with their theoretical values. For the two configurations without active predecessor sites, the activation probabilities remain close to zero across the entire range of $p$. For the four configurations with one active predecessor site, the activation probabilities increase approximately linearly with $p$, whereas the two configurations with two active predecessor sites exhibit a nonlinear dependence consistent with $q=2p-p^2$. The results for all four system sizes closely follow the corresponding theoretical curves, indicating that the network successfully reproduces the local stochastic transition probabilities under the one-step prediction conditions considered here. Moreover, the consistent results across different system sizes demonstrate that the learned local transition rules can be applied to larger systems without retraining.

The preceding results demonstrate that the two-channel convolutional neural network accurately reproduces the one-step activation probabilities of different local configurations. To further assess the performance of the learned stochastic transition rules over multiple time steps, we compare the particle density evolution generated by the neural network with that obtained from direct MC simulations.

A system of size $L=32$ with periodic boundary conditions is initialized in a fully occupied state, with $x_i^0=1$ for all sites. For each propagation probability $p=0.3$, $0.5$, $0.644$, $0.7$, and $0.9$, $2000$ independent stochastic realizations are performed using each method, with the particle density recorded over $t=0$ to $50$. In the ML approach, the trained convolutional neural network predicts the sitewise activation probabilities at each time step, from which the subsequent binary configuration is generated through independent Bernoulli sampling and fed back into the network for further evolution. In contrast, the MC simulations update the configurations directly according to the stochastic propagation rules of directed bond percolation.

\begin{figure}[htbp]
    \centering
    \includegraphics[width=0.95\linewidth]{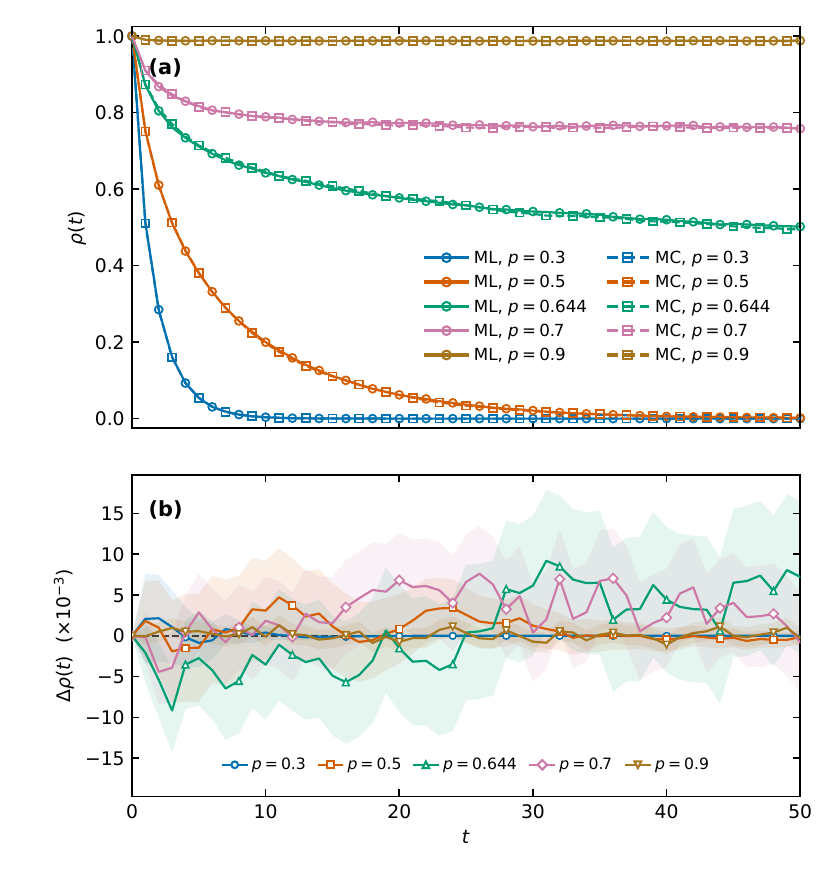}
    \caption{Time evolution of the average particle density at different propagation probabilities and the corresponding differences between ML and MC results. (a) Average particle density $\rho(t)$ as a function of time. Solid lines with open circles represent ML results, while dashed lines with open squares represent MC results. The same color indicates the same propagation probability $p$. (b) Differences in average particle density between ML and MC. The shaded regions represent the approximate pointwise $95\%$ confidence intervals estimated from the standard errors of the independent ensemble means.}
    \label{fig:density_evolution}
\end{figure}

Fig~\ref{fig:density_evolution}(a) compares the temporal evolution of the average particle density obtained from ML and MC at different propagation probabilities. For $p=0.3$ and $0.5$, the particle density decreases rapidly and approaches zero, whereas for $p=0.7$ and $0.9$, the system maintains a relatively high density of active sites over the investigated time interval. Near the critical point, at $p=0.644$, the particle density exhibits a slower decay. The ML-generated density curves closely follow the MC results across the different propagation probabilities, indicating that the learned transition rules reproduce not only the one-step activation probabilities but also the average particle density evolution over multiple time steps.

Fig~\ref{fig:density_evolution}(b) shows the corresponding differences in average particle density, defined as $\Delta\rho(t)=\rho_{\mathrm{ML}}(t)-\rho_{\mathrm{MC}}(t)$. The differences remain relatively small throughout the investigated time interval, although larger fluctuations are observed near the critical region. The shaded regions indicate the approximate pointwise $95\%$ confidence intervals estimated from the standard errors of the independent ensemble means. These results further support the ability of the trained network to reproduce the finite-time density evolution of the DP system.

For a system of size $L=32$ with periodic boundary conditions and a fully occupied initial configuration, we investigate the dependence of the finite-time-averaged particle density on the propagation probability $p$. The maximum evolution time is set to $T=50$, and $51$ equally spaced values of $p$ are selected over the interval $[0,1]$. For each value of $p$, $2000$ independent stochastic realizations are performed using both ML and MC methods.

\begin{figure}[htbp]
    \centering
    \includegraphics[width=0.98\linewidth]{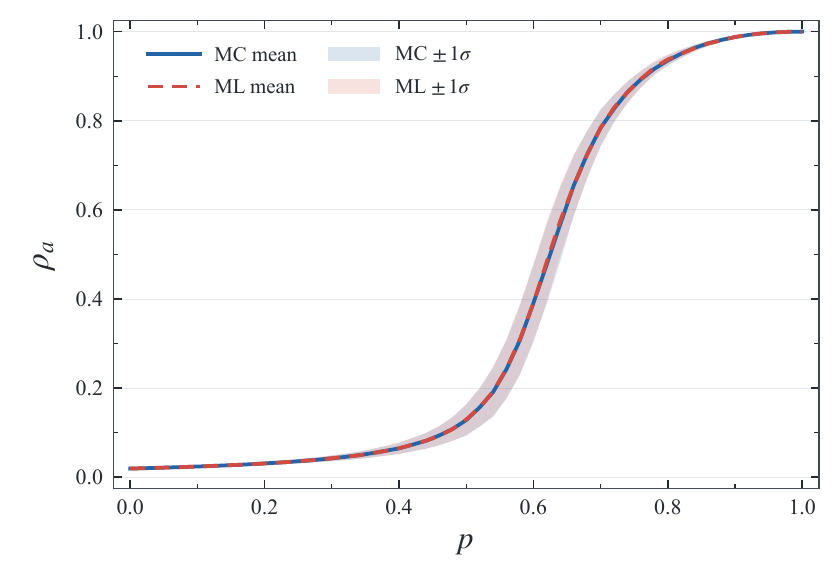}
    \caption{Finite-time-averaged particle density as a function of the propagation probability $p$. The particle density is first averaged over time for each realization and then averaged over $2000$ independent realizations. The blue solid and red dashed lines represent the ensemble averages obtained from MC and ML, respectively. The shaded regions indicate the corresponding means plus or minus one sample standard deviation of the time-averaged densities across independent realizations.}
    \label{fig:density_probability}
\end{figure}

Fig~\ref{fig:density_probability} compares the finite-time-averaged particle densities obtained from ML and MC as a function of $p$. The particle density increases with increasing propagation probability, exhibiting a pronounced rise in the intermediate parameter region. The mean density curves obtained from the two methods nearly overlap across the entire parameter range, while the distributions of time-averaged densities exhibit similar dispersions, as indicated by the shaded regions representing one sample standard deviation. These results further demonstrate that the learned stochastic transition rules reproduce the finite-time-averaged particle density and its statistical variations across different propagation probabilities.

Following the multi-step dynamical tests on finite systems, we further apply the learned stochastic transition rules to a substantially larger system to investigate the long-time density evolution near criticality. The network trained on a system of size $L=32$ is directly applied to larger systems using the same convolutional parameters, without additional training.

\begin{figure}[htbp]
    \centering
    \includegraphics[width=0.98\linewidth]{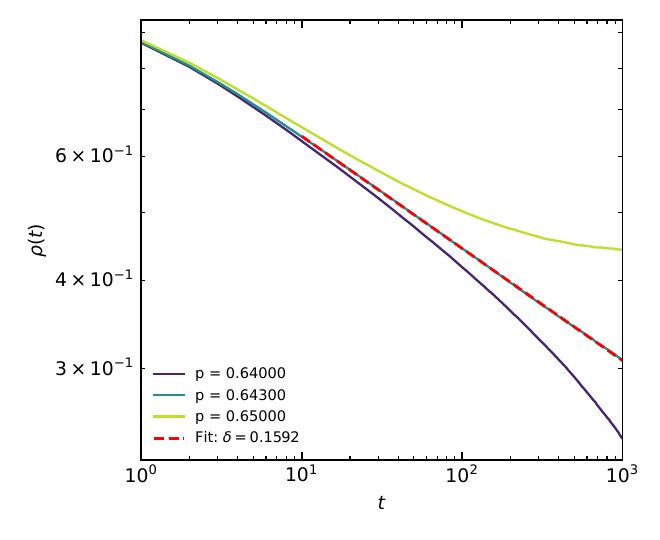}
    \caption{Temporal decay of the ensemble-averaged particle density near criticality using the learned stochastic transition rules. The solid lines represent $\rho(t)$ at different propagation probabilities. The red dashed line indicates the power-law fit $\rho(t)\propto t^{-\delta}$ at $p=0.643$ over the interval $10\leq t\leq1000$, yielding an effective decay exponent of $\delta\approx0.1592$.}
    \label{fig:critical_decay}
\end{figure}

We consider a periodic lattice of size $L=10000$, initialized in a fully occupied state. For propagation probabilities $p=0.640$, $0.643$, and $0.650$, $1000$ independent stochastic realizations are performed at each parameter value, with the evolution extending to $t=1000$. Figure~\ref{fig:critical_decay} shows the temporal evolution of the ensemble-averaged particle density $\rho(t)$ on logarithmic scales, allowing the density decay behavior near the critical region to be examined.

As shown in Fig.~\ref{fig:critical_decay}, the density decay curve for $p=0.640$ bends downward at late times, whereas that for $p=0.650$ gradually approaches a plateau. In contrast, the curve for $p=0.643$ exhibits an approximately linear behavior on the log-log plot, suggesting a power-law decay near criticality. Fitting the density evolution at $p=0.643$ to $\rho(t)\propto t^{-\delta}$ over the interval $10\leq t\leq1000$ yields an effective decay exponent of $\delta\approx0.1592$, which is close to the established value for the (1+1)-dimensional DP universality class. This agreement suggests that the learned local stochastic transition rules can capture the characteristic density decay behavior near criticality, even when applied to systems substantially larger than those used during training.

\section{Conclusion}

In this study, we proposed a stochastic dynamics generative method based on a two-channel convolutional neural network, using DP as a representative model of nonequilibrium phase transitions. Unlike conventional generative approaches that directly model the probability distribution of complete configurations, the proposed method learns local conditional transition probabilities from the current spatial configuration and control parameter, and generates stochastic temporal evolution through Bernoulli sampling. Owing to its fully convolutional architecture, the learned local transition rules can be applied to systems of different spatial sizes without retraining.

We first evaluated the network's ability to learn microscopic stochastic transition rules by examining the one-step activation probabilities of eight local configurations. The predicted probabilities closely follow the theoretical relations $q=0$, $q=p$, and $q=2p-p^2$. Furthermore, the network trained on a system of size $L=32$ reproduces the theoretical activation probabilities for larger systems with $L=64$, $128$, and $256$, demonstrating the applicability of the learned local transition rules across different spatial sizes.

We then investigated the multi-step stochastic evolution generated by the network through comparisons with direct MC simulations. The network reproduces the temporal evolution of the average particle density at different propagation probabilities, including the density decay in the absorbing phase and the sustained activity in the active phase. The finite-time-averaged particle densities also show good agreement with MC results over the investigated parameter range. These findings indicate that the learned local transition probabilities can reproduce the macroscopic statistical behavior of the DP system over multiple time steps.

Finally, we applied the trained network to a substantially larger system with $L=10000$ to investigate the long-time density decay near criticality. At $p=0.643$, the particle density exhibits an approximately power-law decay, yielding a finite-time effective decay exponent of $\delta\approx0.1592$. This result suggests that the learned local stochastic transition rules can capture characteristic density decay behavior near criticality, even in systems considerably larger than those used during training. Overall, this study demonstrates the feasibility of generating nonequilibrium stochastic dynamics by learning local conditional transition probabilities, providing a neural-network-based framework for investigating the microscopic evolution and macroscopic statistical properties of stochastic systems.

\section{Acknowledgements}
This work was supported in part by Yunnan Fundamental Research Projects (Grant NO. 202401AU070035), the Special Basic Cooperative Research Programs of Yunnan Provincial Undergraduate Universities’ Association (grant NO. 202501BA070001-113), and the 111 Project 2.0, with Grant No. BP0820038.

\section{Data sets}
The detailed algorithms of how to generate raw data and implement machine learning are shown in the GitHub link {https://github.com/Gao-DaFeng/Directed-Percolation-ML}.

\bibliographystyle{apsrev4-2}
\bibliography{ref}

\end{document}